\documentclass{IEEEtran}
\usepackage{cite}
\usepackage{amsmath,amssymb,amsfonts}
\usepackage{algorithmic}
\usepackage{graphicx} \graphicspath{{figures/}}
\usepackage{textcomp}
\usepackage{xcolor}
\usepackage{scalerel}
\usepackage{tikz} \usetikzlibrary{svg.path}
\definecolor{orcidlogocol}{HTML}{A6CE39}
\tikzset{
  orcidlogo/.pic={
    \fill[orcidlogocol] svg{M256,128c0,70.7-57.3,128-128,128C57.3,256,0,198.7,0,128C0,57.3,57.3,0,128,0C198.7,0,256,57.3,256,128z};
    \fill[white] svg{M86.3,186.2H70.9V79.1h15.4v48.4V186.2z}
                 svg{M108.9,79.1h41.6c39.6,0,57,28.3,57,53.6c0,27.5-21.5,53.6-56.8,53.6h-41.8V79.1z M124.3,172.4h24.5c34.9,0,42.9-26.5,42.9-39.7c0-21.5-13.7-39.7-43.7-39.7h-23.7V172.4z}
                 svg{M88.7,56.8c0,5.5-4.5,10.1-10.1,10.1c-5.6,0-10.1-4.6-10.1-10.1c0-5.6,4.5-10.1,10.1-10.1C84.2,46.7,88.7,51.3,88.7,56.8z};
  }
}
\newcommand\orcidicon[1]{\href{https://orcid.org/#1}{\mbox{\scalerel*{
\begin{tikzpicture}[yscale=-1,transform shape]
\pic{orcidlogo};
\end{tikzpicture}
}{|}}}}
\usepackage{hyperref}
\def\BibTeX{{\rm B\kern-.05em{\sc i\kern-.025em b}\kern-.08em
    T\kern-.1667em\lower.7ex\hbox{E}\kern-.125emX}}
\begin{document}
\title{Efficient quantification of the impact of demand and weather uncertainty in power system models}
\author{Adriaan P. Hilbers \orcidicon{0000-0002-9882-9479}, \textit{Student Member, IEEE}, David J. Brayshaw \orcidicon{0000-0002-3927-4362}, Axel Gandy \orcidicon{0000-0002-6777-0451}
\thanks{This work was supported by the United Kingdom Engineering and Physical Sciences Research Council (EPSRC) Mathematics of Planet Earth Centre for Doctoral Training, grant number EP/L016613/1. Adriaan Hilbers (e-mail: a.hilbers17@imperial.ac.uk) and Axel Gandy are based at the Department of Mathematics, Imperial College London, London, United Kingdom. David Brayshaw is based at the Department of Meteorology, Reading University, Reading, United Kingdom.}}

\maketitle

\begin{abstract}
This paper introduces a new approach to quantify the impact of forward propagated demand and weather uncertainty on power system planning and operation models. Recent studies indicate that such sampling uncertainty, originating from demand and weather time series inputs, should not be ignored. However, established uncertainty quantification approaches fail in this context due to the data and computing resources required for standard Monte Carlo analysis with disjoint samples. The method introduced here uses an $m$ out of $n$ bootstrap with shorter time series than the original, enhancing computational efficiency and avoiding the need for any additional data. It both quantifies output uncertainty and determines the sample length required for desired confidence levels. Simulations and validation exercises are performed on two capacity expansion planning models and one unit commitment and economic dispatch model. A diagnostic for the validity of estimated uncertainty bounds is discussed. The models, data and code are made available.
\end{abstract}


\section{Introduction}
\label{sec:intro}

\subsection{Demand and weather uncertainty in power system models}
\label{sec:intro_climate_based_uncertainty}

This paper considers \textit{capacity expansion planning} (CEP) and \textit{unit commitment economic dispatch} (UCED) models. Such power system models (PSMs) typically use optimisation to determine e.g. the minimum system cost, optimal generation capacities, generation scheduling or carbon emissions.

Recent studies indicate that the effect of forward propagated \textit{demand and weather uncertainty} on the outputs of such models should not be ignored, especially in systems involving variable renewable generation such as solar and wind \cite{bryce_2018, amorim_2020}. This uncertainty emerges from the use of time series data (e.g. demand levels, wind speeds or solar irradiances) which may be viewed as samples from some underlying demand and weather distribution. This sampling uncertainty may be significant; for example, model outputs may differ highly depending on which year of data is used \cite{bloomfield_2016, staffel_2018, collins_2018, zeyringer_2018, bothwell_2018, kumler_2019}, with some outputs varying as much as 80\% \cite{pfenninger_2017, hilbers_2019}.

\subsection{Uncertainty quantification in power system models}
\label{sec:intro:uncertainty_analysis}

The power system modeling community employs a number of established techniques to quantify output (also called \textit{forward propagated}) uncertainty, as summarised in \cite{soroudi_2013} and \cite{yue_2018}. Scenario analysis is convenient for factors that cannot be reliably described by probability distributions (e.g. future policy or the uptake speed of new technologies), while interval analysis performs well when input parameters lie between certain values and interact weakly. \textit{Monte Carlo} methods estimate the probability distribution of outputs by running a model multiple times with uncertain inputs sampled according to their respective probability distributions; see \cite{pye_2015, alzbutas_2012, bosetti_2015, hart_2011} for some applications to power systems. More sophisticated approaches include creating a statistical emulator for the model \cite{dent_2016}.


\subsection{The m out of n and the time series bootstrap}
\label{sec:intro:bootstrapping}

Since their introduction by Efron \cite{efron_1979}, \textit{bootstrap} methods have become popular throughout statistics and its applications. In its most basic form, the procedure emulates an output's sampling distribution by its distribution under resampling from the available data. Its popularity is attributed to its simplicity and the fact that it ``works'' (referred to as \textit{consistency}) in a wide variety of settings \cite{chernick_book, singh_1981, bickel_1981}. The $m$ out of $n$ bootstrap, which uses samples of a different length (usually shorter), may be used to reduce computational cost. Theoretical properties, including its consistency in settings where the traditional ($n$ out of $n$) bootstrap is consistent, are discussed in \cite{bickel_1997}.

Bootstrap methods applied to time series data require additional refinements when values are not independent and identically distributed (IID), as summarised by \cite{davison_book}. Two common modifications are the sampling of blocks of timesteps to preserve short-term dependence structures (the \textit{block bootstrap}) and detrending of longer-term dependencies and subsequent resampling of residuals (the \textit{model-based bootstrap}).

Bootstrap methods have been applied in a number of energy applications. Some examples include probabilistic forecasts for electricity demand \cite{fan_2012}, price \cite{chen_2012}, and variable renewable generation levels \cite{khosravi_2013, wan_2014}. They have also been used in assessments of system adequacy margins \cite{othman_2008} and, in more physics-oriented studies, electromechanical modes \cite{anderson_2005}.

\subsection{This paper's contribution}
\label{sec:intro:contribution}

This paper provides two contributions. The first is a new scheme to efficiently quantify the impact of demand and weather uncertainty in power systems models (PSMs), determining how much an output varies if other, but equally plausible, demand and weather samples are considered. It is an $m$ out of $n$ bootstrap that uses \textit{shorter} resamples of the available data. The second is a method to determine the required simulation length for desired uncertainty bounds. The methods are applicable to capacity expansion planning and unit commitment economic dispatch models with simulation horizons of at least one year. Both methods are validated experimentally on a wide range of model outputs and a diagnostic for the consistency of estimated uncertainty bounds is found to work well. The models, data and sample code are available at \cite{github_buq}.

Improvements on currently employed approaches (Section \ref{sec:intro:uncertainty_analysis}) are illustrated by an example. Consider uncertainty quantification on PSM outputs obtained using five years of demand and weather data. Scenario and interval analysis fail since the uncertainty originates from high-dimensional time series. Standard Monte Carlo methods require simulations with disjoint five-year samples of demand and weather data \textemdash\, inefficient, both in data (multiple five-year samples) and computation (multiple five-year runs). The method introduced here uses simulations across \textit{shorter} time series (requiring reduced computational resources) resampled from the available data (hence requiring no additional samples). In this way, it reduces the inefficiency of standard Monte Carlo analysis, while maintaining its advantages. For example, the method makes no assumptions on the model formulation and is identical irrespective of the time series inputs, regional topology, or constraints. To the best knowledge of the authors, this paper is the first in the power and energy community that (1) uses shorter samples, reducing the computational requirements of standard bootstrap/Monte Carlo methods, (2) relates sample size to uncertainty levels, allowing an informed choice of simulation length, and (3) quantifies the impact of forward propagated demand and weather uncertainty on a large range of PSM outputs; hitherto, studies focused instead on producing improved point estimates under this uncertainty via e.g. stochastic or robust optimisation or concentrated on only one specific output.

This paper is structured as follows. Section \ref{sec:methodology} introduces the general form of the method. Section \ref{sec:results} analyses its performance on three test power system models: two capacity expansion planning models and one operation (unit commitment economic dispatch) model. Section \ref{sec:discussion_conclusions} discusses the results and conclusions. A full description of the time series and models can be found in the appendix (Section \ref{sec:appendix}).

\section{Methods}
\label{sec:methodology}

\subsection{Overview}
\label{sec:methodology:overview}

Consider a power system model (PSM) output $\hat{O}$ determined using $\mathcal{T}_{\hat{O}}$, a demand and weather time series of length $n_{\hat{O}}$. The PSM is viewed as a mapping from $\mathcal{T}_{\hat{O}}$ to $\hat{O}$:
\begin{equation}
  \label{eq:methodology:output_full}
  \hat{O} = \text{PSM}(\mathcal{T}_{\hat{O}}).
\end{equation}
Uncertainty in the value of $\hat{O}$ is induced from the time series used to calculate it. In particular, a different sample of demand and weather data, drawn from the same underlying distribution as $\mathcal{T}_{\hat{O}}$ is drawn from, leads to a different but equally valid output. This uncertainty on $\hat{O}$, forward propagated from $\mathcal{T}_{\hat{O}}$, is referred to as \textit{demand and weather uncertainty} in this paper.

A concrete example of demand and weather uncertainty is as follows. Consider the cost-optimal level of additional wind capacity to build in order to meet environmental constraints. This depends on the distribution of demand and weather events expected to occur. The optimum additional wind capacity over the last $n_{\hat{O}}$=1 year of demand and weather data provides a point estimate $\hat{O}$ of this optimum. However, a different year-long sample (e.g. the year before) provides a different value. This variation across samples drawn from the same underlying distribution drives \textit{demand and weather uncertainty} on $\hat{O}$.

A convenient quantification of this uncertainty is provided by $\hat{O}$'s sampling standard deviation $\sigma_{\hat{O}}$ across different demand and weather samples, of length $n_{\hat{O}}$, drawn from the same underlying distribution (e.g. 1 year samples in the above example). $\sigma_{\hat{O}}$ can be used to construct confidence or prediction intervals. For example, $[\hat{O} - 2 \sigma_{\hat{O}}, \hat{O} + 2 \sigma_{\hat{O}}]$ has a coverage probability of at least 75\% (and usually higher, e.g. 95\% for a normal distribution). As discussed in Section \ref{sec:intro:contribution}, $\sigma_{\hat{O}}$ often cannot be estimated by Monte Carlo methods with disjoint samples due to limitations in data and computational resources. It may be estimated instead using the $m$ out of $n$ bootstrap as follows:
\begin{enumerate}
\item Construct $K$ subsamples $\mathcal{S}_1, \ldots, \mathcal{S}_K \subset \mathcal{T}_{\hat{O}}$, each of length $n_S \le n_{\hat{O}}$, using a suitable bootstrapping procedure.
\item For each sample, calculate output $O_{k} = \text{PSM}(\mathcal{S}_k)$. Normalise if necessary (see discussion below).
\item Estimate variance $\widehat{\sigma_S}^2$ across subsamples:
  \begin{equation}
    \label{eq:methodology:variance_sample}
    \widehat{\sigma_S}^2 = \frac{1}{K-1} \sum_{k=1}^K (O_{k} - \overline{O_{S}})^2  
  \end{equation}
  where $\overline{O_S} = \frac{1}{K} \sum_{k=1}^K O_{k}$ is the sample mean.
\item Estimate $\sigma_{\hat{O}}$ by $\widehat{\sigma_{\hat{O}}}$, defined by
  \begin{equation}
    \label{eq:methodology:extrapolation}
    {\widehat{\sigma_{\hat{O}}}^2} = \frac{n_S}{n_{\hat{O}}} \widehat{\sigma_S}^2.
  \end{equation}  
\end{enumerate}

The procedure in step 1 works by subsampling, with replacement, blocks of time steps correctly distributed throughout the year. An example is the sampling of weeks from each season. Precise details (e.g. the block length) may be tailored to the specific application as in Section \ref{sec:results:subsampling}. 

The sample length adjustment necessitates the normalisation of $O_k$ in step 2 to lie on the same temporal scale as $\hat{O}$. In this investigation, extensive outputs (e.g. generation levels, emissions and costs) are annualised  and expressed as values per year. Extensive quantities can be restored by multiplying $\hat{O}$ and $\widehat{\sigma_{\hat{O}}}$ by the sample length $n_{\hat{O}}$ at the end. Furthermore, time series should be detrended; for example, long-term demand trends should be removed before resampling.

The method also allows the estimation of the required sample length $n_{\hat{O}}$ to attain desired levels of certainty. Given a required standard deviation $\sigma_{\hat{O}}$, (\ref{eq:methodology:extrapolation}) may be rearranged to give
  \begin{equation}
    \label{eq:methodology:sample_length}
    n_{\hat{O}} = n_S \frac{\widehat{\sigma_S}^2}{\widehat{\sigma_{\hat{O}}}^2}.
  \end{equation}
This motivates the following approach: (1) Use steps 1-3 to obtain $\widehat{\sigma_S}^2$. (2) Estimate $n_{\hat{O}}$, the minimum sample length required to ensure model output $\hat{O}$ has standard deviation at most $\widehat{\sigma_{\hat{O}}}$, using equation (\ref{eq:methodology:sample_length}).

\subsection{Justification and a diagnostic for validity}
\label{sec:methodology:justification_diagnostic}

The standard deviation estimate $\widehat{\sigma_{\hat{O}}}$ is calculated using an $m$ out of $n$ time series bootstrap with $n=n_{\hat{O}}, m=n_S$. From bootstrap theory (Section \ref{sec:intro:bootstrapping}), $\widehat{\sigma_{\hat{O}}}$ is known to approximate $\sigma_{\hat{O}}$ in many settings provided that $n_S$ and $n_{\hat{O}}$ are large enough and the sampled blocks are long enough to capture the autocorrelation in the time series. The $\frac{n_S}{n_{\hat{O}}}$ scaling appears because the variance of many statistical estimators (including functions of the sample mean and median under weak conditions) are inversely proportional to sample size \cite{bickel_1997, van_der_vaart_book}.

Bootstrap theory implies two important cases in which this method is \textit{not} expected to provide consistent estimates. The first is when a PSM output depends on a sample minimum or maximum, e.g. peak net demand or peaking capacity if 100\% of demand must be met. This can be alleviated by allowing unmet demand at high cost, which changes the dependence to a high quantile, e.g.\ 99.9\%. The second is when the PSM output is an integer variable with jumps too large to be approximated well by a continuous one. 

Given a PSM output, a diagnostic gives users an approximate indication whether the method is consistent. It works as follows: use varying subsample length $n_S$ and check whether the estimate $\widehat{\sigma_{\hat{O}}}^2$ in Equation (\ref{eq:methodology:sample_length}) is roughly unchanged. The diagnostic is necessary but not sufficient; the standard deviation estimates are inconsistent if they clearly differ across sample lengths, but constant standard deviation estimates do not prove the consistency of the bootstrap. In practice, the diagnostic is found to work well, succesfully identifying the settings where the standard deviation estimates are consistent across the case studies in this paper.

\section{Simulation results}
\label{sec:results}

\subsection{Overview}
\label{sec:results:overview}

\begin{figure}
  \footnotesize
  \setlength{\tabcolsep}{0.2em}
  \begin{tabular}{c}
    \includegraphics[scale=0.64, trim=10 10 10 10, clip]{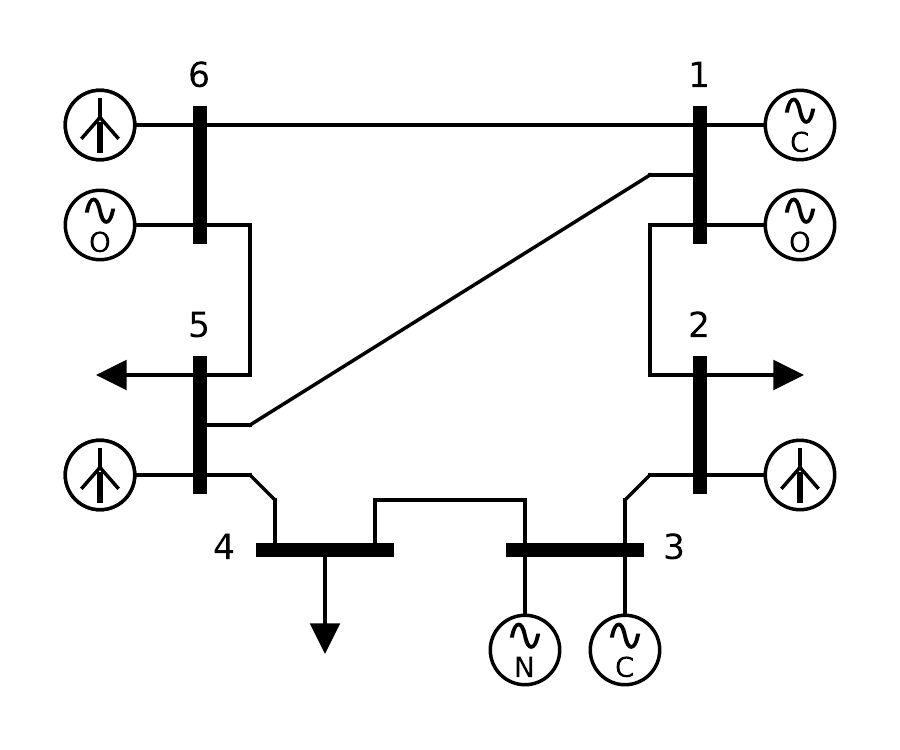}
  \end{tabular}
  \begin{tabular}{c  c}
  Bus & Demand \& generation \\ \hline
  1 & CCGT, OCGT \\
  2 & Demand, Wind (DE) \\
  3 & Nuclear, CCGT \\
  4 & Demand (FR) \\
  5 & Demand, Wind (UK) \\
  6 & OCGT, Wind (ES) \\ \hline
\end{tabular}
  \caption{6-bus model configuration. Demand must be met at buses 2, 5 and 6. CCGT, OCGT and nuclear generation is distributed at buses 1, 3 and 6. Wind generation is available at buses 2, 5 and 6. Buses 2, 4, 5 and 6 use (demand or wind) time series data from Germany (DE), France (FR), the United Kingdom (UK) and Spain (ES) respectively.}
  \label{fig:results:model_diagram}
\end{figure}

In this section the proposed methods are applied to three sample PSMs. Each model's topology is based on the \textit{IEEE 6-bus system} (see e.g.\ \cite{rau_1994, roh_2009, baharvandi_2018}). The available technologies at each bus are based on a renewables-ready version introduced in \cite{kamalinia_2010} and \cite{kamalinia_2011}. Fig. \ref{fig:results:model_diagram} provides a diagram of the model topology and locations of different demand and generation sources. Available generation technologies are nuclear, combined cycle gas turbine (CCGT), open cycle gas turbine (OCGT) and wind. Unmet demand (load shedding) is permitted at high cost. The models' time series inputs are demand levels and wind capacity factors in European countries. The individual models are discussed below, and their technical details can be found in the appendix (Section \ref{sec:appendix}). They are modified versions of a larger class of open-source models and data, available at \cite{github_renewable_test_psms}.

\subsubsection{LP planning model}
\label{sec:results:overview:lp_planning}

a capacity expansion planning model that determines the least cost ``build from scratch'' system design (capacities of generation and transmission technologies) by minimising the sum of install and generation costs. Each generation and transmission capacity may take any nonnegative value. For this reason, the associated optimisation problem is a continuous linear program (LP). Model outputs consist of the cost-optimal generation and transmission capacities as well as the (annualised) minimum system cost.

\subsubsection{MILP planning model}
\label{sec:results:overview:milp_planing}

identical to the \textit{LP planning} model, but with nuclear generation capacity restricted to blocks of 3GW. The optimisation problem is hence a mixed integer linear program (MILP).

\subsubsection{Operation model}
\label{sec:results:overview:operations}

a unit commitment and economic dispatch model that optimises the operation of a system with fixed generation and transmission capacities. Operational constraints are more sophisticated than in the planning models; nuclear power has a ramp rate of 20\%/hr and a minimum operating rate (when switched on) of 50\%. Model outputs consist of generation and transmission levels, generation costs and carbon emissions, all of which are annualised.

The remainder of this section is structured as follows. Section \ref{sec:results:subsampling} introduces the subsampling schemes. Section \ref{sec:results:estimating_standard deviation} discusses the quantification of output uncertainty and Section \ref{sec:results:required_sample_length} illustrates how to estimate the required sample length for desired confidence levels. The method's validity is examined in Section \ref{sec:results:verification}.

\subsection{Subsampling schemes}
\label{sec:results:subsampling}

This section describes how the $K$ bootstrap samples (step 1, Section \ref{sec:methodology:overview}) are created. Subsampling schemes should avoid distorting power system operation (and hence PSM outputs) when compared to the original time series. For example, for ramping constraints or storage, the chronology of time steps should be altered as little as possible, and when seasonal demand or wind patterns exist, the distribution of time steps throughout the year should be preserved. For this reason, two stratified block subsampling schemes are proposed:
\begin{itemize}
\item \textbf{Months}: sample individual months, correctly distributed throughout the year. For example, a two-year sample is:
\begin{equation}
  \small
  \label{eq:sample_months}
  [\text{Jan}][\text{Feb}] \cdots [\text{Nov}][\text{Dec}] [\text{Jan}][\text{Feb}] \cdots [\text{Nov}][\text{Dec}]
\end{equation}
where e.g. $[\text{Jan}]$ is a contiguous January block sampled from the original time series $\mathcal{T}_{\hat{O}}$.
\item \textbf{Weeks}: sample individual weeks, correctly distributed throughout the year. For example, a 56-day sample is:
\begin{equation}
  \label{eq:sample_weeks}
  [\text{7d}]_{_\text{DJF}} [\text{7d}]_{_\text{MAM}} [\text{7d}]_{_\text{JJA}} [\text{7d}]_{_\text{SON}} [\text{7d}]_{_\text{DJF}} [\text{7d}]_{_\text{MAM}} [\text{7d}]_{_\text{JJA}} [\text{7d}]_{_\text{SON}}
\end{equation}
where e.g. $[\text{7d}]_{_\text{DJF}}$ is a contiguous week block from one of the meteorological winters (Dec-Jan-Feb) in $\mathcal{T}_{\hat{O}}$.
\end{itemize}

\subsection{Quantifying output uncertainty}
\label{sec:results:estimating_standard deviation}

\begin{table}
  \caption{Details on the simulations in Section \ref{sec:results:estimating_standard deviation}. The mean values per simulation are shown.}
  \centering
  \footnotesize
  \setlength{\tabcolsep}{0.3em}
  \begin{tabular}{r | c | c | c}
    & (a) & (b) & (c) \\
    Model & LP planning & MILP planning & operation \\
    Sample for point estimate $\hat{O} $& 2008-2017 & 2017 & 2017 \\
    Sample length $n_{\hat{O}} $& 10yr & 1yr & 1yr \vspace*{1.5em} \\
    \multicolumn{4}{c}{\textbf{Standard deviation estimate via $m$ out of $n$ bootstrap}} \\ \hline
    Subsample scheme & \textit{months} & \textit{weeks} & \textit{months} \\
    Sample length $n_S$ & 1yr & 12wk & 1yr \\
    Mean solution time (minutes) & 6 & 14 & 5 \\
    Mean solution memory (MB) & 1633 & 457 & 332 \vspace*{1.5em} \\
    \multicolumn{4}{c}{\textbf{Standard deviation estimate via Monte Carlo with disjoint samples}} \\ \hline
    Sample length & 10yr & 1yr & 1yr \\
    Mean solution time (minutes) & 262 & 765 & 5 \\
    Mean solution memory (MB) & 13813 & 1713 & 332 \\
  \end{tabular}
  \label{tab:example}
\end{table}

\begin{figure}
  \center{(a) \textit{LP planning} model, point estimate 2008-2017} \vspace*{0.1em} \\
  \includegraphics[scale=0.57, trim=10 10 0 5, clip]{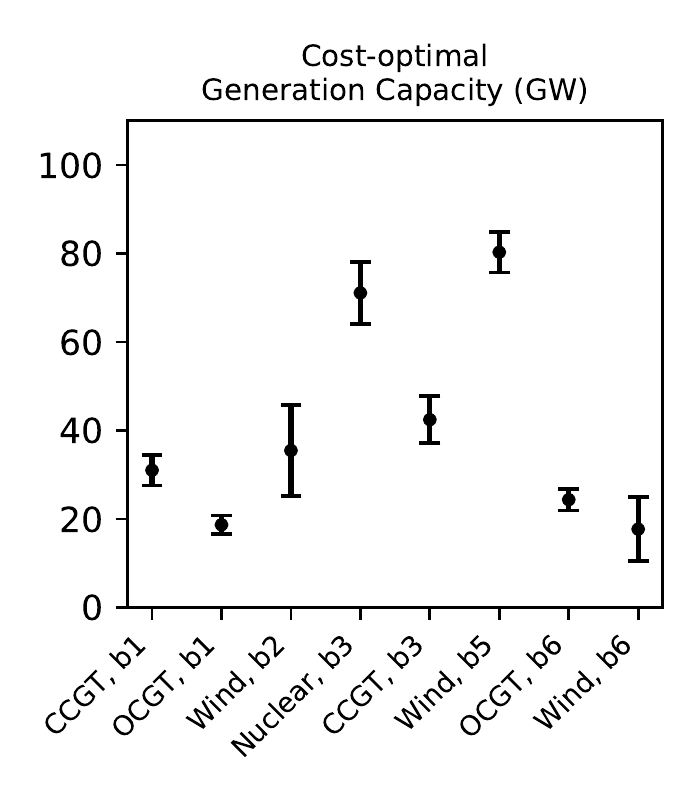}
  \includegraphics[scale=0.57, trim=10 10 0 5, clip]{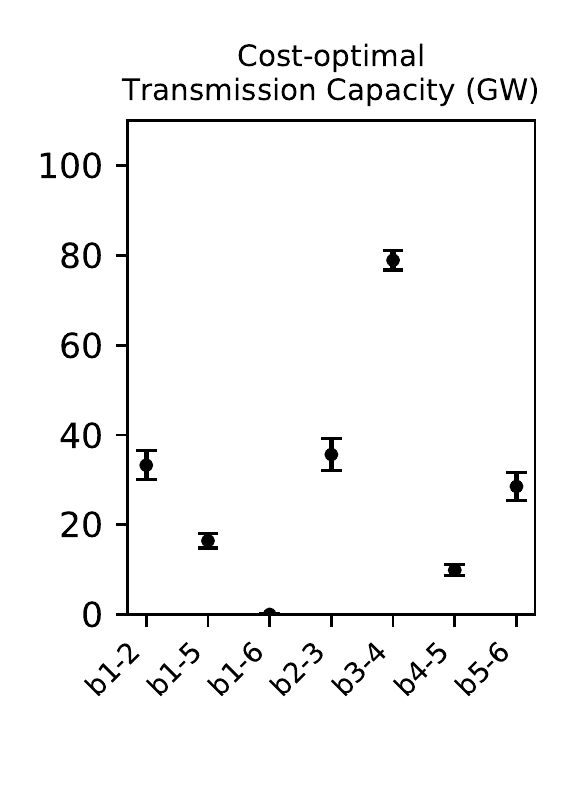}
  \includegraphics[scale=0.57, trim=10 10 10 5, clip]{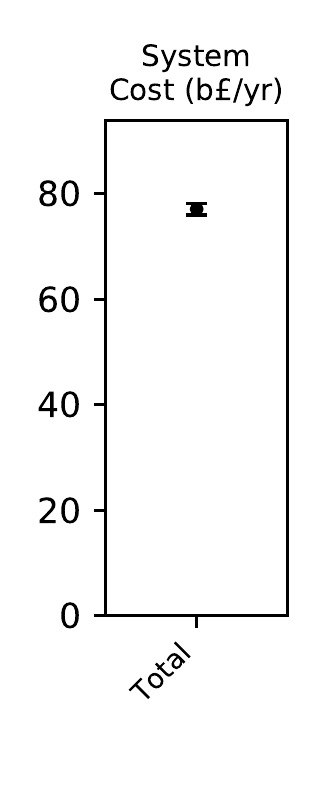} \\ \vspace*{0.5em}
  \center{(b) \textit{MILP planning} model, point estimate 2017} \vspace*{0.5em} \\
  \includegraphics[scale=0.57, trim=10 10 0 5, clip]{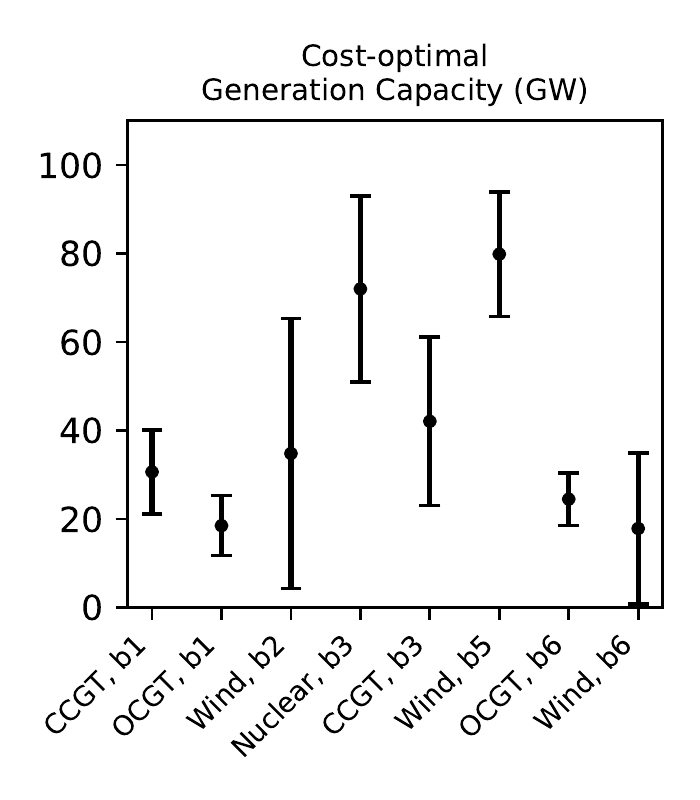}
  \includegraphics[scale=0.57, trim=10 10 0 5, clip]{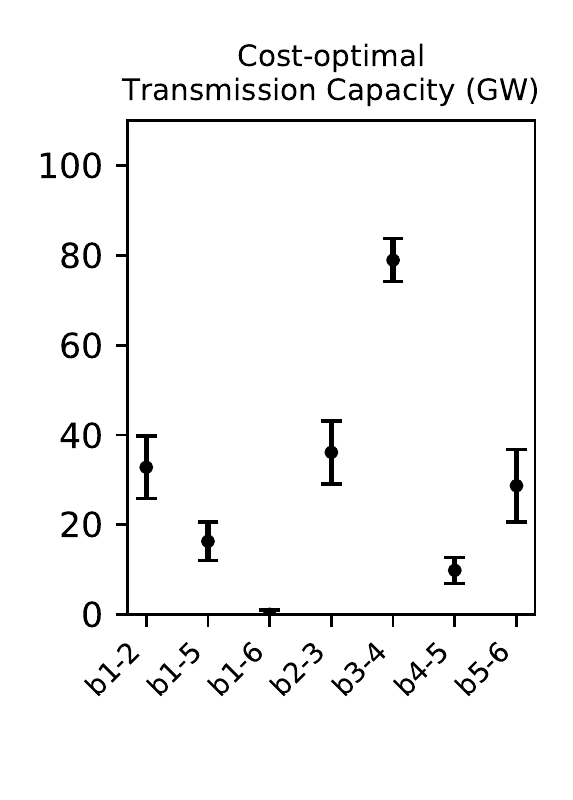}
  \includegraphics[scale=0.57, trim=10 10 10 5, clip]{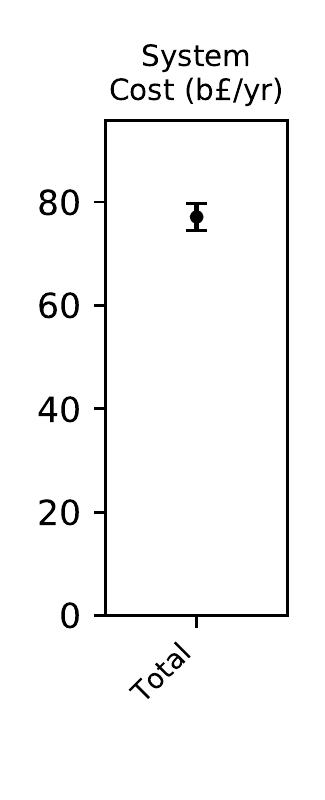} \\ \vspace*{0.5em}
  \center{(c) \textit{Operation} model, point estimate 2017} \vspace*{0.5em} \\
  \includegraphics[scale=0.57, trim=10 10 5 5, clip]{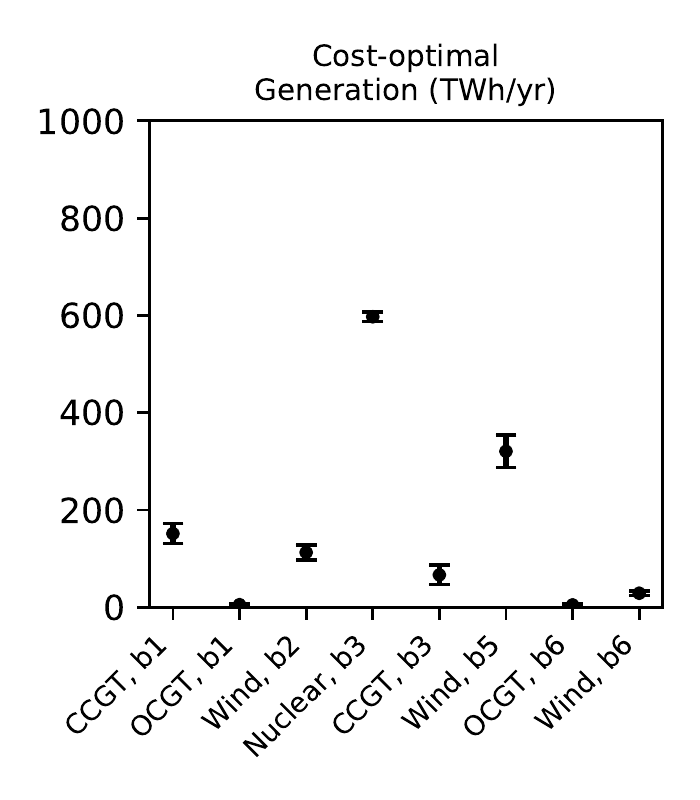}
  \includegraphics[scale=0.57, trim=10 10 5 5, clip]{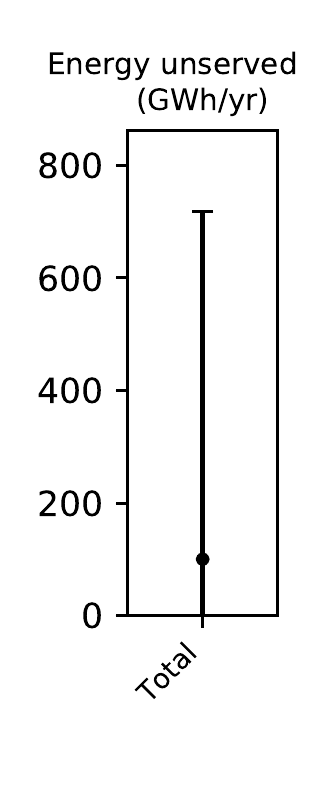}
  \includegraphics[scale=0.57, trim=10 10 0 5, clip]{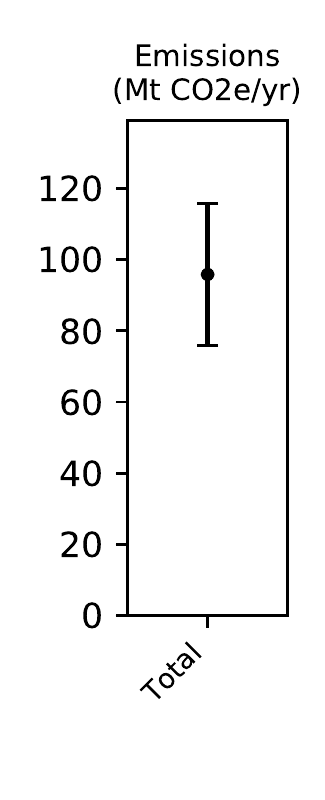}
  \includegraphics[scale=0.57, trim=10 10 10 5, clip]{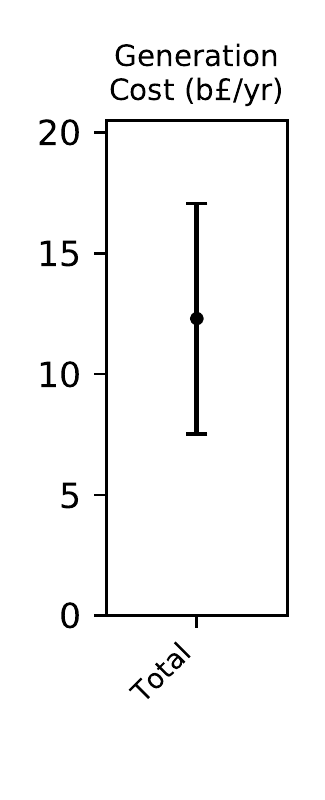}
  \caption{Point estimates of selected model outputs, with error bars equal to 2 standard deviations. The standard deviation is estimated using the $m$ out of $n$ bootstrap with $K$=1,000 bootstrap samples. In the $x$-axis labels, ``bX'' means bus number X. See Table \ref{tab:example} for details on the simulations.}
  \label{fig:model_outputs}
\end{figure}

In this section, the demand and weather uncertainty on model outputs from each of the three sample PSMs is quantified. Let each (point estimate) model output be $\hat{O}$, determined using a single long simulation. Estimates for $\hat{O}$'s sampling standard deviation $\sigma_{\hat{O}}$ are generated using the $m$ out of $n$ bootstrap (Section \ref{sec:methodology:overview}) with $K$=1,000 bootstrap samples. The time series that is sampled from is the period 2008-2017.

Table \ref{tab:example} details the simulations employed. It also provides, for comparison, the simulations required for standard Monte Carlo methods using disjoint samples of the same length as the point estimate (the current approach). Optimisation problems are created using the energy modeling framework \textsl{Calliope} \cite{pfenninger_2018} and solved using the \textsl{CBC} optimiser \cite{cbc} on a 2.7GHz Intel Core i5-5257U processor with 8GB of RAM (with additional ``swap'' memory).

One test case is run for each PSM. In test case (a), the \textit{LP planning} model is considered. The point estimate $\hat{O}$ is calculated across 2008-2017 and the standard deviation estimate $\widehat{\sigma_{\hat{O}}}$ uses the $m$ out of $n$ bootstrap with subsample length $n_S$=1 year, generated by resampling months as per the \textit{months} scheme (Section \ref{sec:results:subsampling}). In test case (b), the \textit{MILP planning} model is run across 2017 for the point estimate and standard deviation estimates use subsample length $n_S$=12 weeks, generated by resampling weeks from seasons as per the \textit{weeks} scheme (Section \ref{sec:results:subsampling}). In test case (c), which considers the \textit{operation} model, the sample lengths for the point estimate and bootstrap simulations are equal. This corresponds to a ``regular'' ($n$ out of $n$) time series bootstrap.

Fig. \ref{fig:model_outputs} shows, for a subselection of model outputs, the point estimates $\hat{O}$ with error bars of length $2\widehat{\sigma_{\hat{O}}}$ above and below. The same plots for the full range of model outputs are available in the supplementary material. This range typically covers at least 75\% of a distribution, and typically more (95\% for a normal distribution). In Fig. \ref{fig:model_outputs}(a), the use of a 10-year sample for the point estimate means uncertainty levels are relatively small. Cost-optimal wind capacity at bus 2 has a $\hat{O} \pm 2\widehat{\sigma_{\hat{O}}}$ range of 25-45GW, but other outputs are comparatively certain. For Fig. \ref{fig:model_outputs}(b), the use of a shorter (1-year) point estimate means uncertainty bounds are much larger. For example, $\hat{O} \pm 2\widehat{\sigma_{\hat{O}}}$ extends from 2-62GW for optimal wind capacity at bus 2. Furthermore, in this case, a user could not say with any certainty whether, for example, cost-optimal system design has more CCGT capacity at bus 1 or 3, as the uncertainty regions overlap considerably. In test case (a), in contrast, there is a high confidence that more CCGT capacity should be installed at bus 3. In both settings (a) and (b), optimal transmission capacities and minimum system cost are comparatively certain. In Fig. \ref{fig:model_outputs}(c), the uncertainty is large for energy unserved but much smaller for the generation levels. The total generation cost is also relatively uncertain, ranging roughly 40\% above and below. This is driven almost totally by the uncertainty in unserved energy, which is assigned a high value of lost load and hence a high contribution to total costs.

The $m$ out of $n$ bootstrap estimates the standad deviation significantly more efficiently (both in data and computation) than standard Monte Carlo methods with disjoint samples and without a reduction in sample length. For example, in test case (a), the method represents a 44-fold reduction in computing time (6 vs. 262 minutes) and an eight-fold reduction in memory (1633 vs. 13813 MB). Since the estimates are generated across $K$=1,000 samples, the total computing time in this case is 6,000 minutes (100 hours), compared with 262,000 ($\approx$182 days) using standard Monte Carlo methods. Furthermore, the resampling means 1,000 bootstrap samples can be created without any additional data. For the same number of samples using standard Monte Carlo methods, one requires 1,000 samples of ten years each: 10,000 years of data in total. Similar advantages occur in test case (b). In test case (c), the standard deviation estimates are calculated without a reduction in sample length, so that $n_S = n_{\hat{O}}$. In this case, the advantages are only in data efficiency; instead of requiring 1,000 years of data, bootstrap samples can be created by resampling the original ten years.

\subsection{Determining required sample length}
\label{sec:results:required_sample_length}

The relationship between sample size and standard deviation allows an informed choice of simulation length based on desired certainty levels. For example, suppose a user wants $\widehat{\sigma_{\hat{O}}}$ on optimal nuclear capacity at bus 3 to be no more than $\sigma_{\hat{O}}$=5GW. In test case (a), the method indicates that $\sigma_S$, the standard deviation across 1,000 $n_S$=1-year bootstrap samples, is about 11GW. Using equation (\ref{eq:methodology:sample_length}) then indicates a required sample length of $n_{\hat{O}} = 1 \times \frac{11^2}{5^2} \approx 5$ years for the point estimate. Calculations for other outputs work in the same way.

\subsection{Verifying the method's consistency}
\label{sec:results:verification}

In this section, the method's consistency is considered. This is done by using the $m$ out of $n$ bootstrap, with varying subsample length $n_S$, to estimate $\sigma_Y$, the sampling standard deviation of single-year model outputs. A 95\% confidence interval for $\sigma_Y$ is constructed by bootstrapping the 38 individual (disjoint) years from 1980-2017, a well-established procedure \cite{van_der_vaart_book}. If the method is consistent, each standard deviation estimate meets two conditions: (1) it lies within this confidence interval with high probability, and (2) it is independent of the sample length $n_S$ used to calculate it (indicating that the $\frac{n_S}{n_{\hat{O}}}$ scaling factor used in step 4 of Section \ref{sec:methodology:overview} is appropriate). 

\begin{figure}
  \center{(a) \textit{LP planning model}} \vspace{0.5em} \\
  \includegraphics[scale=0.57, trim=10 40 10 22, clip]{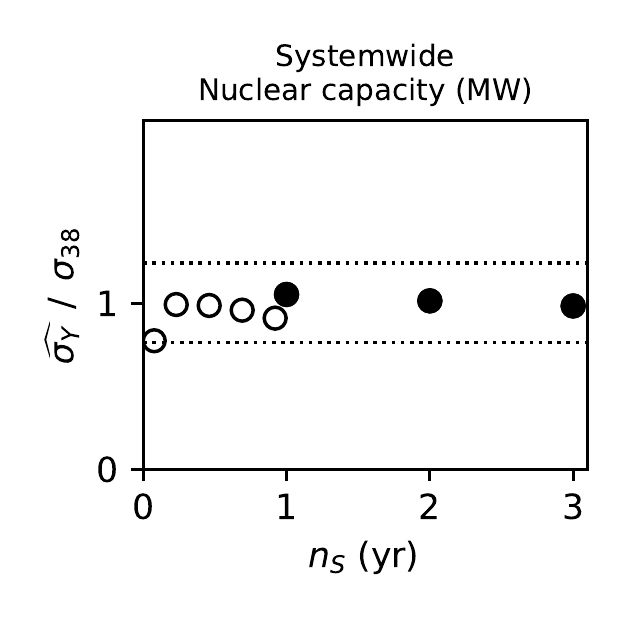}
  \includegraphics[scale=0.57, trim=38 40 10 22, clip]{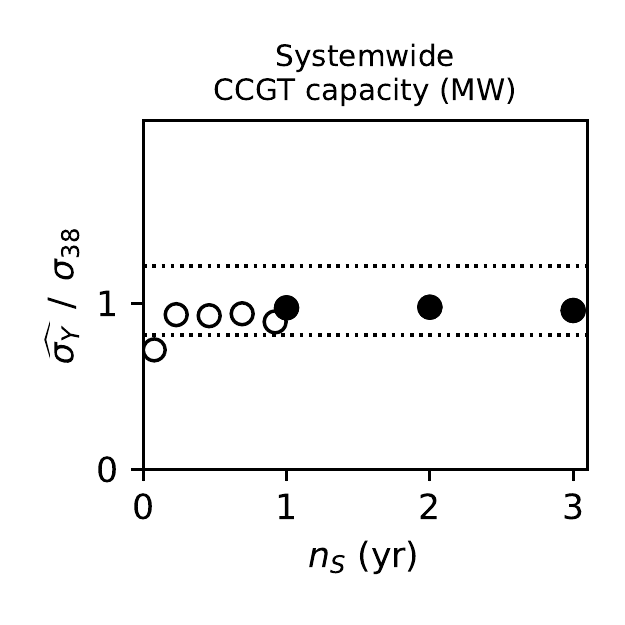}
  \includegraphics[scale=0.57, trim=38 40 10 22, clip]{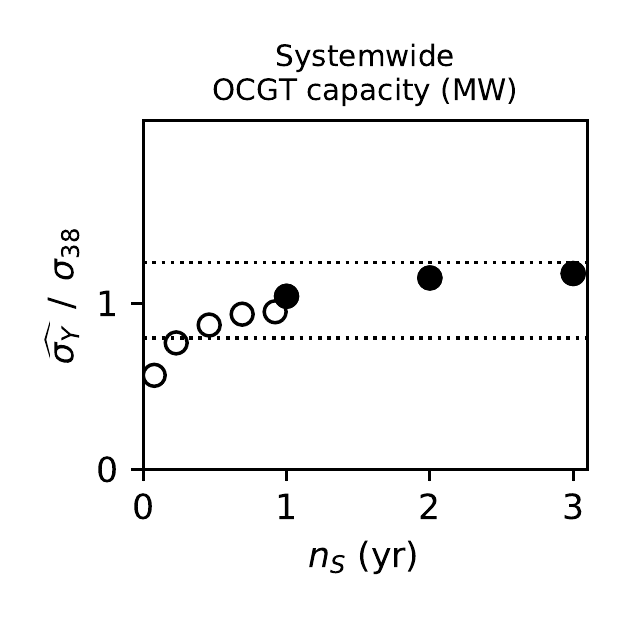} \\ \vspace*{0.5em}
  \includegraphics[scale=0.57, trim=10 40 10 22, clip]{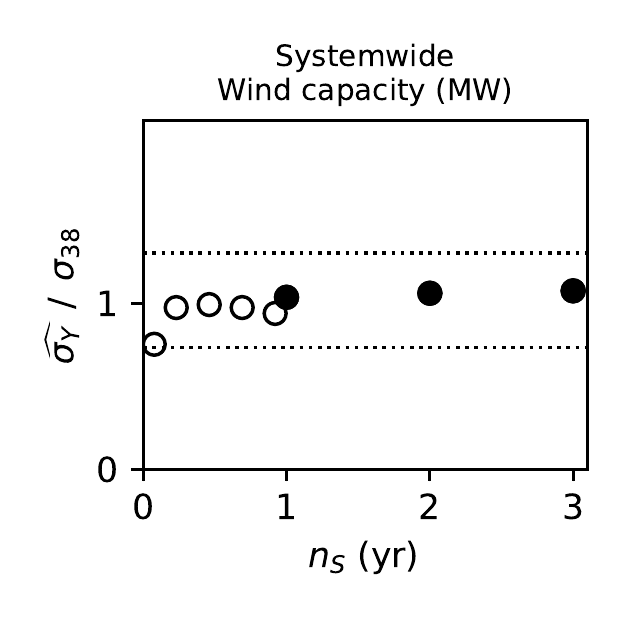}
  \includegraphics[scale=0.57, trim=38 40 10 22, clip]{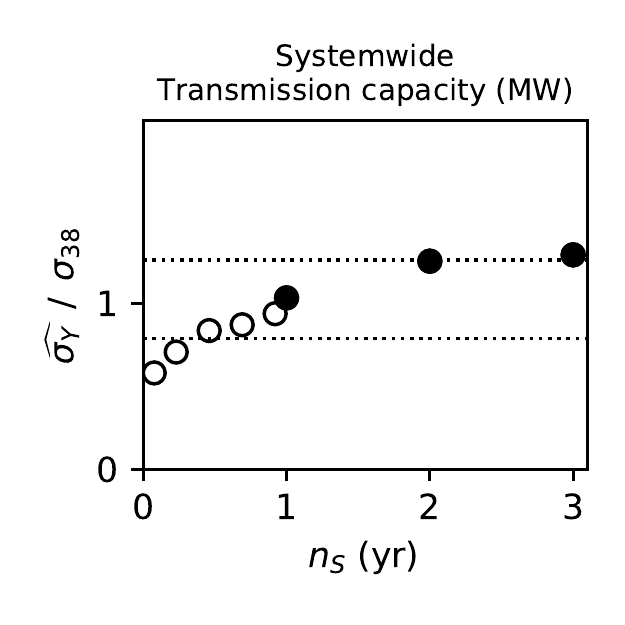}
  \includegraphics[scale=0.57, trim=38 40 10 22, clip]{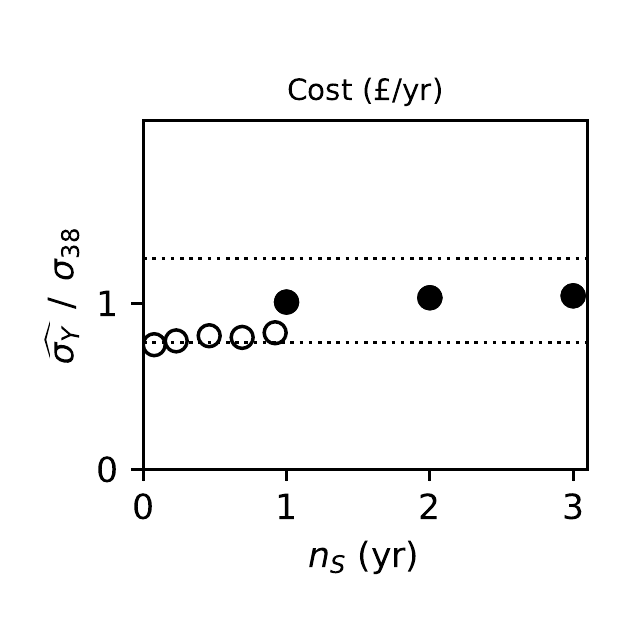}

  \center{(b) \textit{MILP planning model}} \vspace{0.5em} \\
  \includegraphics[scale=0.57, trim=10 40 10 22, clip]{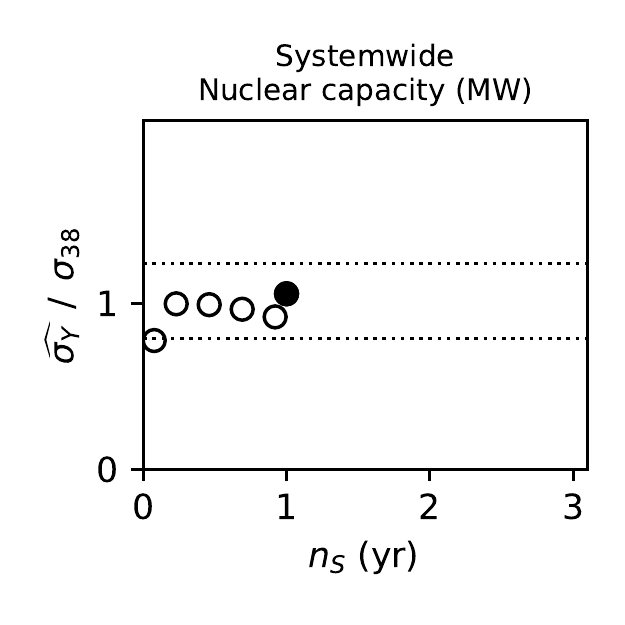}
  \includegraphics[scale=0.57, trim=38 40 10 22, clip]{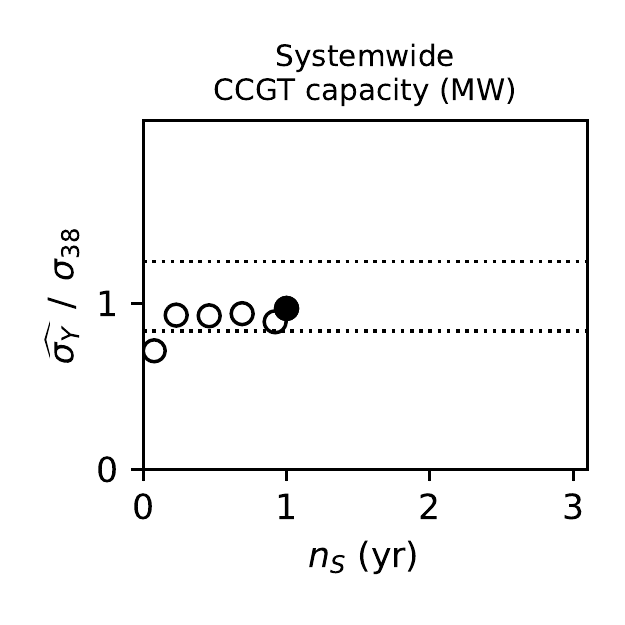}
  \includegraphics[scale=0.57, trim=38 40 10 22, clip]{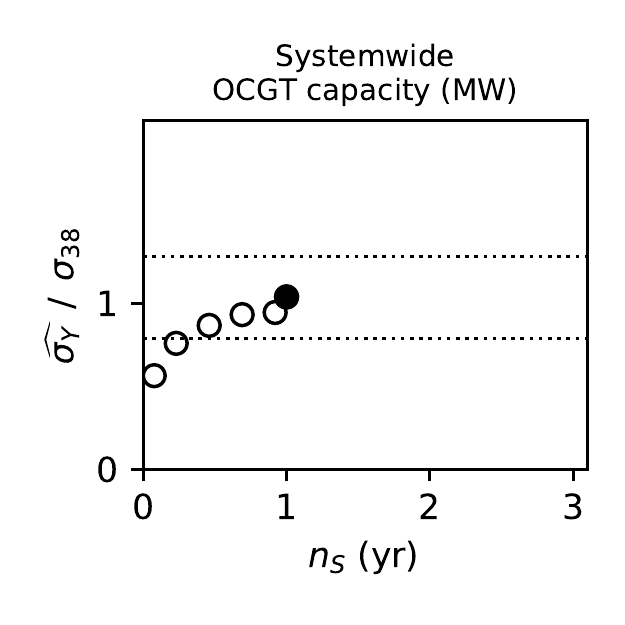} \\ \vspace*{0.5em}
  \includegraphics[scale=0.57, trim=10 40 10 22, clip]{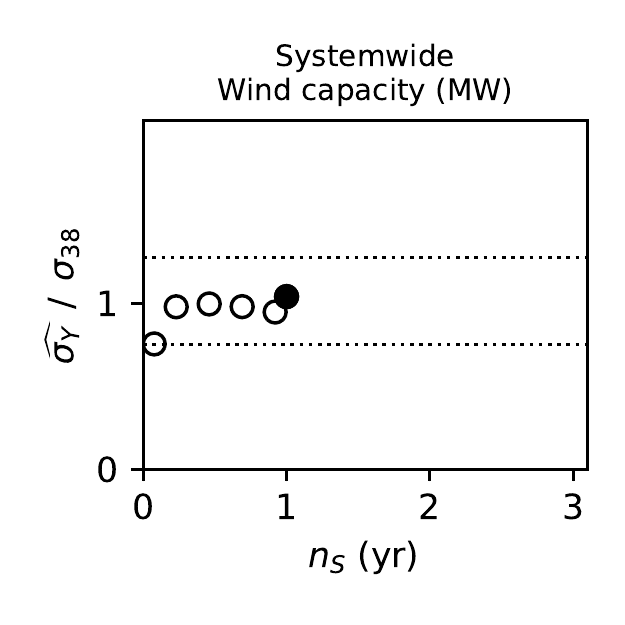}
  \includegraphics[scale=0.57, trim=38 40 10 22, clip]{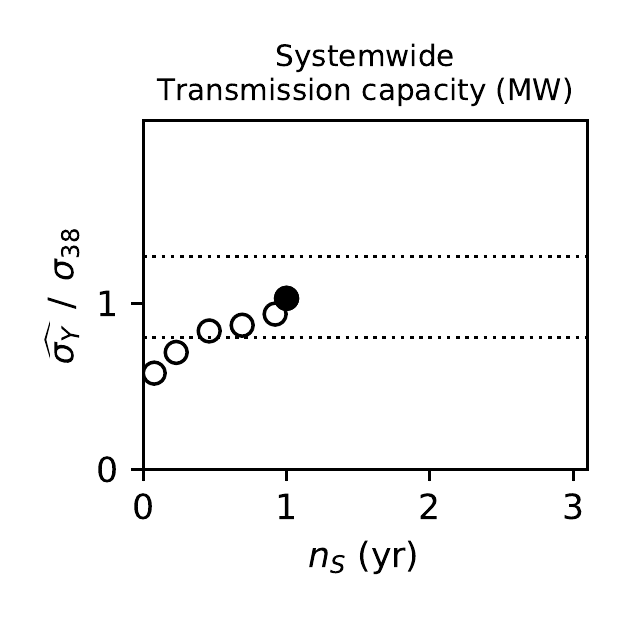}
  \includegraphics[scale=0.57, trim=38 40 10 22, clip]{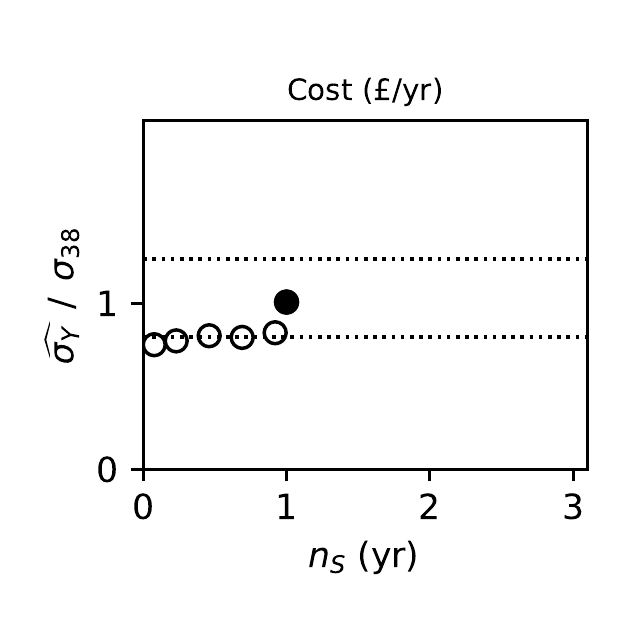}

  \center{(c) \textit{Operation model}} \vspace{0.5em} \\
  \includegraphics[scale=0.57, trim=10 40 10 22, clip]{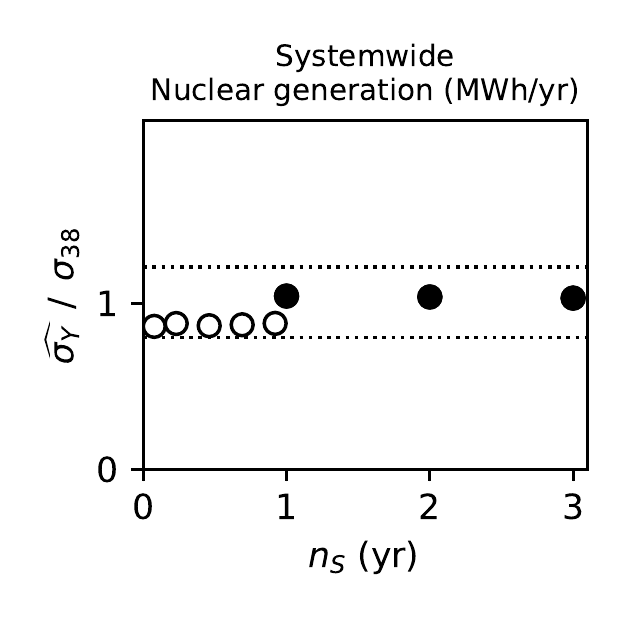}
  \includegraphics[scale=0.57, trim=38 40 10 22, clip]{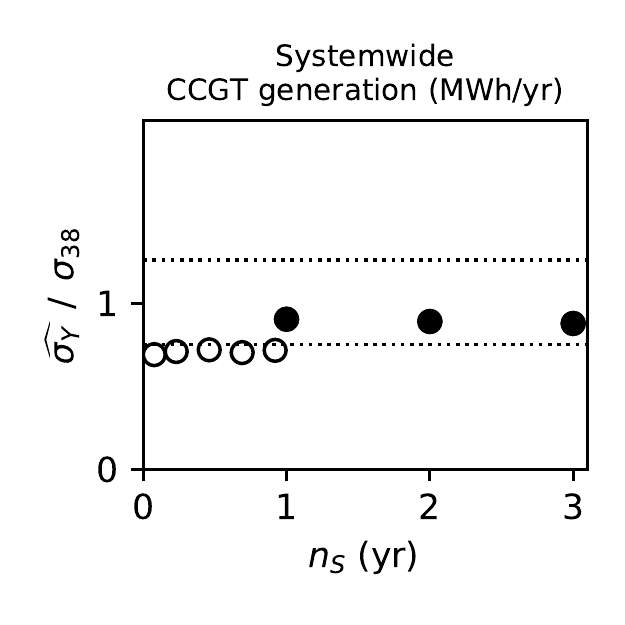}
  \includegraphics[scale=0.57, trim=38 40 10 22, clip]{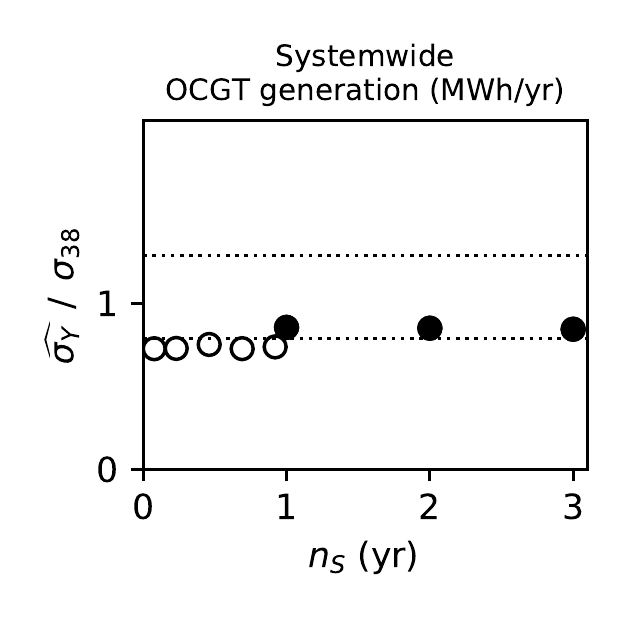} \\ \vspace*{0.5em}
  \includegraphics[scale=0.57, trim=10 40 10 22, clip]{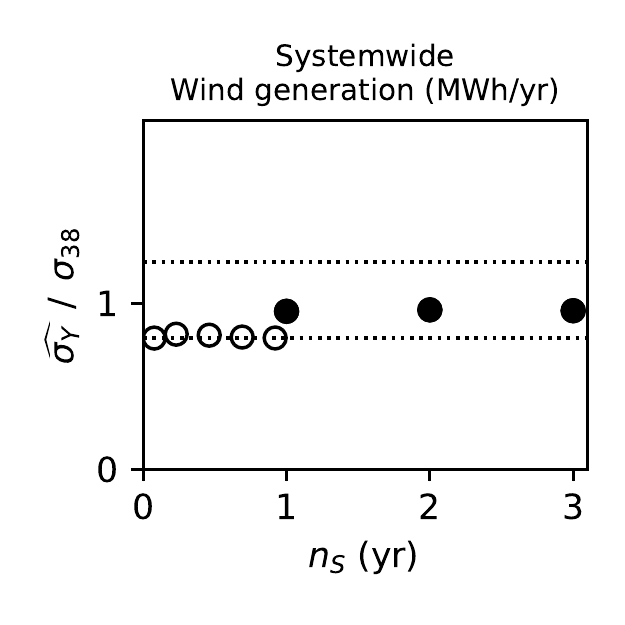}
  \includegraphics[scale=0.57, trim=38 40 10 22, clip]{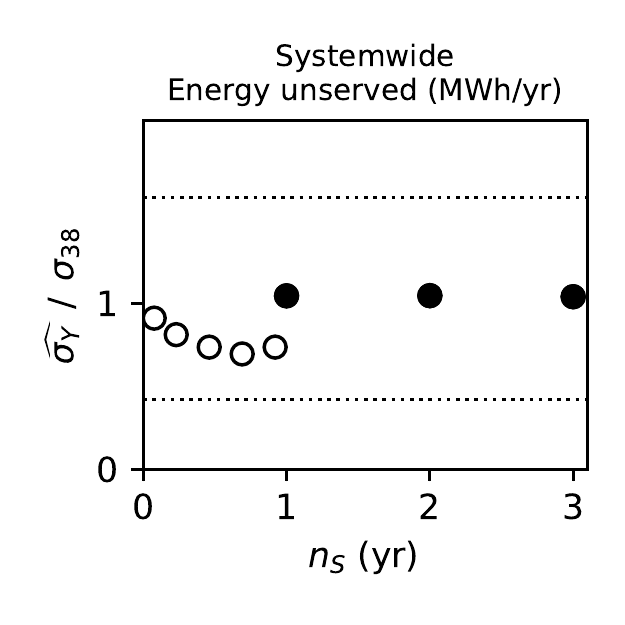}
  \includegraphics[scale=0.57, trim=38 40 10 22, clip]{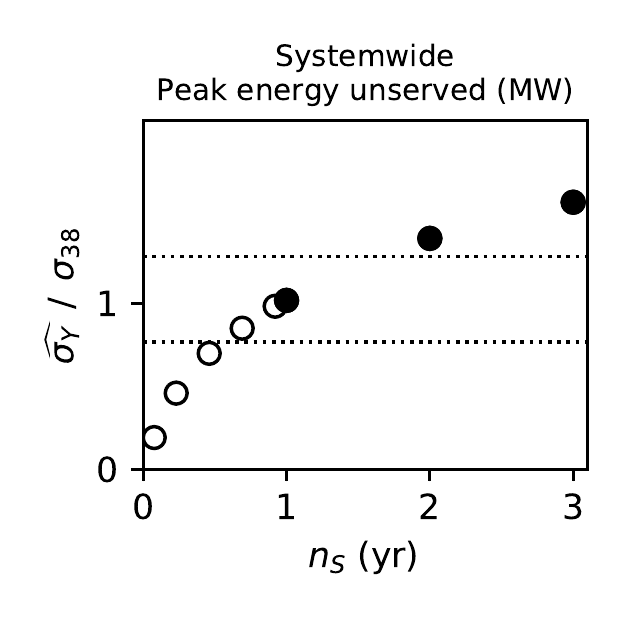} \\ \vspace*{0.5em}
  \includegraphics[scale=0.57, trim=10 40 10 22, clip]{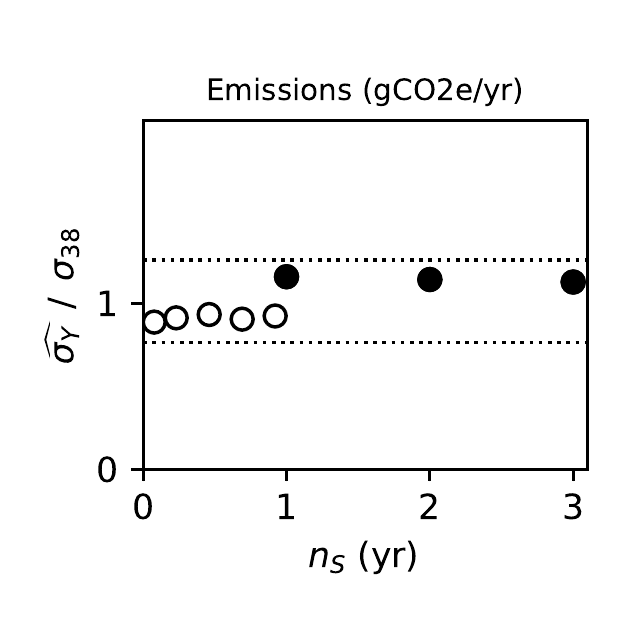}
  \includegraphics[scale=0.57, trim=38 40 10 22, clip]{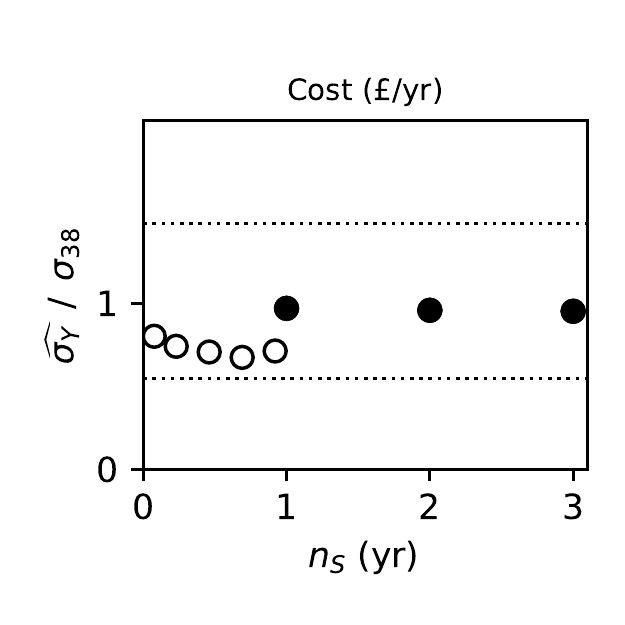} \hspace*{7.6em} \\ \vspace*{2.0em}
  \hspace*{1.2em}
  \includegraphics[scale=0.57, trim=38 10 10 130, clip]{6_region_operate/figures/sd_scaling/gen_wind_total.pdf}
  \includegraphics[scale=0.57, trim=38 10 10 130, clip]{6_region_operate/figures/sd_scaling/gen_wind_total.pdf}
  \includegraphics[scale=0.57, trim=38 10 10 130, clip]{6_region_operate/figures/sd_scaling/gen_wind_total.pdf} \\

  \hspace*{1em} \includegraphics[scale=0.55, trim=0 0 0 0, clip]{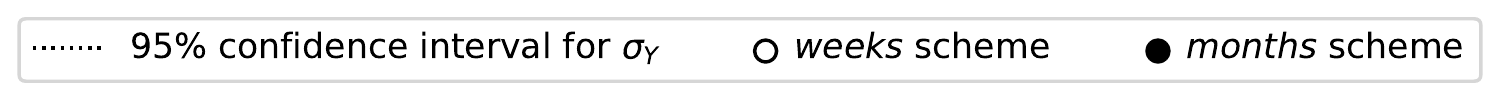} \\
  \caption{(\textit{All plots share the same $x$-axis, shown at the bottom.}) Standard deviation estimates $\widehat{\sigma_Y}$ as a function of subsample size $n_S$. $\widehat{\sigma_Y}$ estimates $\sigma_Y$, the sampling standard deviation of single-year model outputs. $\widehat{\sigma_Y}$ is calculated across $K$=1,000 bootstrap samples, generated by either the \textit{months} or \textit{weeks} scheme. The dashed lines show a symmetric 95\% confidence interval for $\sigma_Y$. The $y$-values are shown as a proportion of $\sigma_{38}$, the standard deviation across the 38 individual years from 1980-2017.}
  \label{fig:sd_estimates_sample_size}
\end{figure}

Fig. \ref{fig:sd_estimates_sample_size} shows, for a selection of model outputs, the estimates $\widehat{\sigma_Y}$ along with a 95\% confidence interval for the true value $\sigma_Y$. Plots for the full range of model outputs are available as supplementary material. Estimates using subsamples shorter than 1 year ($n_S$=4, 12, 24, 36 and 48 weeks) are generated using the \textit{weeks} scheme, while subsamples at least one year ($n_s$=1, 2 and 3 years) are generated via the \textit{months} scheme. For the \textit{operation} model with $n_s\in \{2, 3\}$, the bootstrap samples are longer than the sample used to calculate the point estimate $\hat{O}$ in Section \ref{sec:results:estimating_standard deviation}. This is only to validate the method across a large range of subsample sizes $n_S$; in any practical application, bootstrap samples would be no longer than that for the point estimate. For the \textit{MILP planning} model, the longest sample length is 1 year since the integer constraints, and associated scaling of computational cost with sample length, precludes the use of longer samples.

Fig. \ref{fig:sd_estimates_sample_size} indicates that, with sufficient subsample size, the standard deviation estimates for almost all model outputs meet the two required conditions; they lie within the 95\% confidence interval and are roughly independent of sample length (as seen by a constant $y$-value). The one output where estimates are clearly not consistent is for peak energy unserved in Fig. \ref{fig:sd_estimates_sample_size}(c). The reason for this, as discussed in Section \ref{sec:methodology:justification_diagnostic}, is that this model output is a sample maximum, namely that of demand minus available generation.

The approximate diagnostic introduced in Section \ref{sec:methodology:justification_diagnostic} succesfully identifies for which outputs the $m$ out of $n$ bootstrap does not provide consistent estimates. This diagnostic stipulates that the estimate $\widehat{\sigma_Y}$ is roughly independent of subsample size $n_S$. All outputs in Fig. \ref{fig:sd_estimates_sample_size} except peak energy unserved in Fig. \ref{fig:sd_estimates_sample_size}(c) meet this requirement. These are precisely the model outputs that consistently lie within (or very close to) the confidence interval.

Two further observations warrant mention. The first is that the method fails when subsample sizes are too small, as for transmission capacity with $n_S$=4 weeks in Figures \ref{fig:sd_estimates_sample_size}(a)-(b). The second is that the \textit{weeks} scheme usually generates slightly lower standard deviation estimates than the \textit{months} scheme and may underestimate the target $\sigma_Y$, as in Fig. \ref{fig:sd_estimates_sample_size}(c). This likely occurs because there is some dependence at time scales longer than seven days which is treated as zero when resampling, artificially reducing the standard deviation. The error induced by this approach is not very large, rarely exceeding 15\%. Furthermore, the independent resampling of blocks from different years does not seem to lead to major underestimation of variance by ``smoothing'' over inter-year variability as described in \cite{bodini_2016}.

\section{Discussion and conclusions}
\label{sec:discussion_conclusions}

\subsection{Discussion}
\label{sec:discussion_conclusions:discussion}

The simulation study highlights a number of key messages. The first is that the method usually ``works'', providing uncertainty estimates at a greatly reduced computational cost and without requiring any additional data. The method fails when the subsample size is too small (as is virtually always the case in computational uncertainty quantification) or when the model output depends strongly on a sample maximum (e.g. maximum demand). Each of these failures are identified by the diagnostic introduced at the end of Section \ref{sec:methodology:justification_diagnostic}, giving users an indication whether estimates are consistent. Furthermore, the method provides consistent estimates for nuclear capacity for the \textit{MILP planning} model, even though this is a discrete variable in blocks of 3GW. This is because the 3GW jumps are small enough to make this discrete variable well approximated by a continuous one.

This paper also illustrates, in accordance with previous studies (Section \ref{sec:intro_climate_based_uncertainty}), the risk in informing energy strategy on short demand and weather samples. For example, the cost optimal system design for the \textit{LP planning} model contains 15GW of wind capacity at bus 2 under 1982 data, but 92GW under 1986 data, a more than sixfold difference. Hence, for robust PSM outputs under demand and weather uncertainty, longer samples, spanning multiple years, should be considered.

The discussion above leads to three recommendations. The first is to make the subsamples used for the point estimate as long as possible, as this reduces the uncertainty on their values. The second is that, if both are computationally feasible, the \textit{months} scheme should be preferred over the \textit{weeks} scheme, as the former has less tendency to underestimate variability. The third is to estimate the standard deviation with multiple subsample lengths and check if uncertainty estimates are close, since a failure for this diagnostic indicates that estimates are inconsistent.

\subsection{Conclusion}
\label{sec:discussion_conclusion:conclusion}

This paper introduces a new approach to efficiently quantify the impact of forward propagated demand and weather uncertainty in power system models. This provides an information advantage over point estimates by indicating whether an output is statistically robust or an artefact of the particular demand and weather sample. The method is applicable to both planning (capacity expansion) and operation (unit commitment and economic dispatch) models with simulation lengths of at least one year and without large seasonal storage capacities. Furthermore, simulation lengths may be informed based on desired uncertainty levels, avoiding needless computational expense. The method can be expected to ``work'' (generate consistent uncertainty estimates) for most PSM outputs, and an approximate diagnostic can help determine (but not prove) whether this is the case. The models, data and example code are available at \cite{github_buq}.

There are a number of possible extensions to this study. One is to combine the proposed methods with uncertainty analysis on other model outputs (e.g. technology or fuel costs) by sampling both demand and weather data and other uncertain input parameters simultaneously. Another possible extension is the analysis of the \textit{covariance} of model outputs instead of, as in this investigation, viewing each as having an independent distribution. A third extension involves modifying the method for models with seasonal storage. Finally, the method could leverage the use of subsampling for point estimates. Due to computational limitations, even point estimate PSM outputs are often determined using a subsampled time series such as a smaller number of ``representative days'' \cite{hoffmann_2020, hilbers_2020_iss}. Such approaches can be straightforwardly combined with the algorithm proposed in this paper by using subsampled data for the point estimate. However, more sophisticated methods, such as using subsampled data in the bootstrap samples, or accounting for the uncertainty induced by subsampling for the point estimate, require additional study.

\section{Appendix}
\label{sec:appendix}

\begin{table}
  \caption{Nomenclature.}
  \centering
  \footnotesize
  \setlength{\tabcolsep}{0.2em}
\begin{tabular}{r p{3.6cm}}
  Term & Description \\ \hline
  \multicolumn{2}{l}{\textbf{Indices / Sets}} \\
  $i$ & Generation technology \\
  $r$ & Region (bus) \\
  $t$ & Time step \\
  $\mathcal{I}$ & Technologies: nuclear ($n$), CCGT ($c$), OCGT ($o$), wind ($w$), unmet demand ($u$) \\
  $\mathcal{R}$ & Regions (buses): 1-6 \\
  \multicolumn{2}{l}{\textbf{Parameters}} \\
  $C_{i}^\text{gen}$ & Annualised install cost, technology $i$ (\pounds/MWyr)\\
  $C_{r, r'\!}^\text{tr}$ & Annualised install cost, transmission, bus $r$ to $r'$ (\pounds/MWyr) \\
  $F_{i}^\text{gen}$ & Generation cost, technology $i$ (\pounds/MWh) \\ \hline
\end{tabular} \hspace{0.1em}
\begin{tabular}{r p{3.0cm}}
  Term & Description \\ \hline
  \multicolumn{2}{l}{\textbf{Time series}} \\
  $T$ & Simulation length (hr) \\
  $d_{r, t}$ & Demand, region $r$, time $t$ (MWh) \\
  $w_{r, t}$ & Wind capacity factor, bus $r$, time $t$ ($\in$ [0, 1]) \\
  \multicolumn{2}{l}{\textbf{Variables}} \\
  $\text{cap}_{i, r}^\text{gen}$ & Generation capacity, technology $i$, bus $r$ (MW) \\
  $\text{cap}_{r, r'\!}^\text{tr}$ & Transmission capacity, bus $r$ to $r'$ (MW) \\
  $\text{gen}_{i, r, t}$ & Generation, technology $i$, bus $r$, time $t$ (MWh) \\
  $\text{tr}_{r, r', t}$ & Transmission, bus $r$ to $r'$, time $t$ (MWh) \\ \hline
\end{tabular}
\label{table:appendix:nomenclature}
\end{table}

\subsection{Mathematical formulation of power system models}
\label{sec:appendix:optimisation}

\subsubsection{LP planning model}
\label{sec:appendix:optimisation_LP}

Model outputs are determined by solving (\ref{eq:model_3:objective})-(\ref{eq:model_3:ge_0}) \textbf{without} necessarily satisfying (\ref{eq:model_3:integer}).

\subsubsection{MILP planning model}
\label{sec:appendix:optimisation_MILP}

This model determines its outputs by solving the following optimisation problem:

\vspace{0em}
\begin{scriptsize}
\begin{equation}
\min \! \sum_{r \in \mathcal{R}} \Bigg[ \frac{T}{8760} \Bigg( \underbrace{\sum_{i \in \mathcal{I}} C_i^\text{gen} \text{cap}_{i \!,\! r}^\text{gen}}_{\substack{\text{installation cost,} \\ \text{generation capacity}}} + \underbrace{ \frac{1}{2} \! \sum_{r' \! \in \! \mathcal{R}} \! C_{r \!,\! r'}^\text{tr} \text{cap}_{r \!,\! r'}^\text{tr} }_{\substack{\text{installation cost,} \\ \text{transmission capacity}}} \Bigg) + \underbrace{ \sum_{i \in \mathcal{I}} \sum_{t=1}^{T} F_i^\text{gen} \text{gen}_{i \!,\! r \!,\! t}}_\text{generation cost} \Bigg]
\label{eq:model_3:objective}
\end{equation}
by optimising over decision variables
\begin{equation}
\{\text{cap}_{i\!,r}^\text{gen}, \text{cap}_{r\!,r\!'\!}^\text{tr}, \text{gen}_{i\!,r\!,t}, \text{tr}_{r\!,r\!'\!\!,t} \hspace{0.5em} : \hspace{0.5em} i \!\in\! \mathcal{I}; \hspace{0.5em} r,r' \!\in\! \mathcal{R}; \hspace{0.5em} t \!\in\! \{1 \ldots T \} \}
\label{eq:model_3:decision_variables}
\end{equation}
\noindent subject to
\begin{align}
\text{cap}_{n \!, r}^\text{gen} \!\big\rvert_{r \notin \! \{\! 1 \!\}} \!\!= \text{cap}_{c \!, r}^\text{gen} \!\big\rvert_{r \notin \! \{\! 1 \!,\! 3 \!\}} \!\!= \text{cap}_{o \!, r}^\text{gen} \!\big\rvert_{r \notin \! \{\! 1 \!,\! 6 \!\}} \!\!= \text{cap}_{w \!, r}^\text{gen} \!\big\rvert_{r \notin \! \{\! 2 \!,\! 5 \!,\! 6 \!\}} \!\! &= 0 \label{eq:model_3:topology_gen} \\
\text{cap}_{r \!,\! r'}^\text{tr} \!\big\rvert_{(\! r \!, r' \!) \notin \! \{\! (\!1 \!,\! 2\!) \!,\! (\!1 \!,\! 5\!) \!,\! (\!1 \!,\! 6\!) \!,\! (\!2 \!,\! 1\!) \!,\! (\!2 \!,\! 3\!) \!,\! (\!3 \!,\! 2\!) \!,\! (\!3 \!,\! 4\!) \!,\! (\!4 \!,\! 3\!) \!,\! (\!4\!,\! 5\!) \!,\! (\!5 \!,\! 1\!) \!,\! (\!5 \!,\! 4\!) \!,\! (\!5 \!,\! 6\!) \!,\! (\!6 \!,\! 1\!) \!,\! (\!6 \!,\! 5\!) \!\}} &= 0 \label{eq:model_3:topology_tr}\\
\sum_{i \in \mathcal{I}} \text{gen}_{i,r,t} + \sum_{r' \in \mathcal{R}} \text{tr}_{r',r,t} = d_{r,t} \quad & \forall \: r, t \label{eq:model_3:demand_met} \\
\text{tr}_{r,r',t} + \text{tr}_{r,'r,t} = 0 \quad & \forall \: r, r', t \label{eq:model_3:tr_balance} \\
\text{gen}_{i,r,t} \le \text{cap}_{i,r}^\text{gen} \quad & \forall \: i \!\ne\! w, \forall \: r, t \label{eq:model_3:gen_le_cap_conv} \\
\text{gen}_{w,r,t} \le \text{cap}_{w,r}^\text{gen} w_{r,t} \quad & \forall \: r, t \label{eq:model_3:gen_le_cap_wind} \\
|\text{gen}_{n,r,t} - \text{gen}_{n,r,t+1}| \le 0.2 \text{cap}_{n,r}^\text{gen} \quad & \forall \: r, t \label{eq:model_3:ramping} \\
\text{cap}_{n,r}^\text{gen} \in 3\mathbb{Z} \quad & \forall \: r \label{eq:model_3:integer} \\
\text{cap}_{r,r'}^\text{tr} = \text{cap}_{r',r}^\text{tr} \quad & \forall \: r, r' \label{eq:model_3:tr_cap_symmetrical} \\
|\text{tr}_{r,r',t}| \le \text{cap}_{r,r'}^\text{tr} \quad & \forall \: r, r', t \label{eq:model_3:tr_le_cap_tr} \\
\text{cap}_{i,r}^\text{gen}, \text{cap}_{r,r'}^\text{tr}, \text{gen}_{i,r,t} \ge 0 \quad & \forall \: i, r, t. \label{eq:model_3:ge_0}
\end{align}
\end{scriptsize}
\noindent\eqref{eq:model_3:topology_gen}-\eqref{eq:model_3:topology_tr} stipulate the model's generation and transmission topology. \eqref{eq:model_3:demand_met} and \eqref{eq:model_3:tr_balance} are the demand and power flow balance requirements. \eqref{eq:model_3:gen_le_cap_conv}-\eqref{eq:model_3:gen_le_cap_wind} ensure generation does not exceed installed capacity (for thermal technologies) or installed capacity times the wind capacity factor (for wind). \eqref{eq:model_3:ramping} is the nuclear ramping constraint. \eqref{eq:model_3:integer} enforces that nuclear is built in 3GW units. \eqref{eq:model_3:tr_cap_symmetrical} and \eqref{eq:model_3:tr_le_cap_tr} stipulate that transmission capacities are symmetric and limit transmitted power to installed transmission capacity.

\subsubsection{Operation}
\label{sec:appendix:optimisation:operations}

This model has fixed generation and transmission capacities, equal to the cost-optimal system of the \textit{MILP planning} model across the year 2017. It determines its model outputs by solving the following optimisation problem:

\vspace{0em}
\begin{scriptsize}
\begin{equation}
\min \underbrace{ \sum_{r \in \mathcal{R}} \sum_{i \in \mathcal{I}} \sum_{t=1}^{T} F_i^\text{gen} \text{gen}_{i \!,\! r \!,\! t}}_\text{generation cost}
\label{eq:model_4:objective}
\end{equation}
by optimising over decision variables
\begin{equation}
\{ \text{gen}_{i\!,r\!,t}, \text{tr}_{r\!,r\!'\!\!,t} \hspace{0.5em} : \hspace{0.5em} i \!\in\! \mathcal{I}; \hspace{0.5em} r,r' \!\in\! \mathcal{R}; \hspace{0.5em} t \!\in\! \{1 \ldots T \} \}
\label{eq:model_4:decision_variables}
\end{equation}
\noindent subject to
\begin{align}
\sum_{i \in \mathcal{I}} \text{gen}_{i,r,t} + \sum_{r' \in \mathcal{R}} \text{tr}_{r',r,t} = d_{r,t} \quad & \forall \: r, t \label{eq:model_4:demand_met} \\
\text{tr}_{r,r',t} + \text{tr}_{r,'r,t} = 0 \quad & \forall \: r, r', t \label{eq:model_4:tr_balance} \\
\text{gen}_{i,r,t} \le \text{cap}_{i,r}^\text{gen} \quad & \forall \: i \!\ne\! w, \forall \: r, t \label{eq:model_4:gen_le_cap_conv} \\
\text{gen}_{w,r,t} \le \text{cap}_{w,r}^\text{gen} w_{r,t} \quad & \forall \: r, t \label{eq:model_4:gen_le_cap_wind} \\
|\text{gen}_{n,r,t} - \text{gen}_{n,r,t+1}| \le 0.2 \text{cap}_{n,r}^\text{gen} \quad & \forall \: r, t \label{eq:model_4:ramping} \\
\text{gen}_{n,r,t} = 0 \quad \text{OR} \quad \text{gen}_{n,r,t} \ge 0.5\text{cap}_{n,r}^\text{gen} \quad & \forall \: r, t \label{eq:model_4:unit_commitment} \\
|\text{tr}_{r,r',t}| \le \text{cap}_{r,r'}^\text{tr} \quad & \forall \: r, r', t \label{eq:model_4:tr_le_cap_tr} \\
\text{gen}_{i,r,t} \ge 0 \quad & \forall \: i, r, t. \label{eq:model_4:ge_0}
\end{align}
\end{scriptsize}
\eqref{eq:model_4:demand_met} and \eqref{eq:model_4:tr_balance} are the demand and power flow balance requirements. \eqref{eq:model_4:gen_le_cap_conv}-\eqref{eq:model_4:gen_le_cap_wind} ensure generation does not exceed installed capacity (for thermal technologies) or installed capacity times the wind capacity factor (for wind). \eqref{eq:model_4:ramping} is the nuclear ramping constraint. \eqref{eq:model_4:unit_commitment} indicates that nuclear power must generate at 50\% its rated capacity whenever it is turned on. \eqref{eq:model_4:tr_le_cap_tr} limits transmitted power to installed transmission capacity.

\subsection{Technologies and time series}
\label{sec:appendix:techs_time_series}

\begin{table}
\caption{Generation and transmission technologies. Installation costs are annualised to reflect cost per year of lifetime.}
  \centering
  \setlength{\tabcolsep}{0.4em}
\begin{tabular}{ r  c  c  c}
\small
& Installation cost & Generation cost & Carbon emissions\\
Technology &(\pounds/KWyr) &(\pounds/KWh) & (gCO$_2$e/KWh) \\ \hline
\multicolumn{4}{c}{\textbf{Generation}} \\  
Nuclear & $C_b^\text{gen} = 300$ & $F_b^\text{gen} = 0.005$ & $e_b = 200$ \\
CCGT & $C_p^\text{gen} = 100$ & $F_p^\text{gen} = 0.035$ & $e_m = 400$ \\
OCGT & $C_p^\text{gen} = 50$ & $F_p^\text{gen} = 0.100$ & $e_m = 400$ \\
Wind & $C_w^\text{gen} = 100$ & $F_w^\text{gen} = 0$ & $e_w = 0$ \\
Unmet demand & $C_u^\text{gen} = 0$ & $F_u^\text{gen} = 6$ & $e_u = 0$ \\
\multicolumn{4}{c}{\textbf{Transmission}} \\
Bus 1 to 5 & $C_{1,5}^\text{tr} = 150$ & - & - \\
Bus 1 to 6 & $C_{1,5}^\text{tr} = 130$ & - & - \\
Other & $C_{r,r'}^\text{tr} = 100$ & - & - \\ \hline
\end{tabular} \vspace{1em}
\label{table:appendix:tech_characteristics}
\end{table}

\textit{Nuclear, CCGT, OCGT and wind} are based on the \textit{baseload, mid-merit, peaking} and \textit{wind} technologies in \cite{hilbers_2019}. Carbon emissions are based on \cite{electric_insights}. Unmet demand is considered, for modeling purposes, a fourth technology with no installation cost but a generation cost equal to the value of lost load in the UK \cite{elexon_2015}.

The time series are country-aggregated hourly demand levels (with long-term trends removed) and wind capacity factors for different European countries over the period 1980-2017. Details can be found in \cite{bloomfield_2019} and \cite{bloomfield_MERRA2}.

\bibliographystyle{IEEEtran}
\bibliography{citations}

\end{document}